\documentclass[12pt]{article}%
\usepackage{amsmath,latexsym}
\usepackage{graphicx}
\usepackage{amsmath}
\usepackage{amsfonts}
\usepackage{amssymb}
\begin{document}

\title{{A survey of recent studies concerning the
   extreme properties of Morris-Thorne wormholes}}
   \author{
Peter K.F. Kuhfittig*\\  \footnote{kuhfitti@msoe.edu}
 \small Department of Mathematics, Milwaukee School of
Engineering,\\
\small Milwaukee, Wisconsin 53202-3109, USA}

\date{}
 \maketitle

\begin{abstract}\noindent
It has been known for a long time that
Morris-Thorne wormholes can only be held
open by violating the null energy condition,
which can be expressed in the form
$\tau-\rho c^2>0$, where $\tau$ is the
radial tension.  Matter that violates this
condition is usually referred to as
``exotic."  For any wormhole having a
moderately-sized throat, the radial tension
is equal to that at the center of a massive
neutron star.  Attributing this outcome to
exotic matter seems reasonable enough, but
it ignores the fact that exotic matter was
introduced for a completely different reason.
Moreover, its problematical nature  suggests
that the amount of exotic matter be held to
a minimum, but this would make the high
radial tension harder to explain.  If the
amount is infinitely small, this explanation
breaks down entirely.  By invoking $f(R)$
modified gravity, the need for exotic matter
at the throat could actually be eliminated,
but the negation of the above condition, i.e.,
$\tau <\rho  c^2$, shows that we have not
necessarily eliminated the high radial
tension.  This survey discusses various
ways to account for the high radial tension
and, in some cases, the possible origin
of exotic matter.  We conclude with some
comments on a possible multiply-connected
Universe. \\

\end{abstract}

\section{Introduction}\label{S:introduction}
This paper is a brief survey of recent
studies dealing with some of the extreme
aspects of Morris-Thorne wormholes, such
as the existence and possible origin of
exotic matter and the often inexplicably
large radial tension at the throat.  Most
of the discussion is based in Refs.
\cite{pK13a, pK20, pK21a, pK21b}.

\subsection{Morris-Thorne wormholes}\label{S:MT}

Wormholes are handles or tunnels connecting
widely separated regions of our Universe or
different universes in a multiverse.  Apart
from some forerunners, macroscopic traversable
wormholes were first proposed by Morris and
Thorne \cite{MT88} in 1988.  They introduced
the following static and spherically symmetric
line element for a wormhole spacetime:
\begin{equation}\label{E:line1}
  ds^{2}=-e^{\nu(r)}dt^{2}+e^{\lambda(r)}dr^2
  +r^{2}(d\theta^{2}
  +\text{sin}^{2}\theta\,d\phi^{2}),
\end{equation}
where
\begin{equation}\label{E:lambda1}
   e^{\lambda(r)}=\frac{1}
  {1-\frac{b(r)}{r}}.
\end{equation}
(We are using units in which $c=G=1$.)  Here
$\nu=\nu(r)$ is called the \emph{redshift
function}, which must be everywhere finite
to prevent the occurrence of an event
horizon.  The function $b=b(r)$ is called
the \emph{shape function} since it
determines the spatial shape of the
wormhole when viewed, for example, in an
embedding diagram \cite{MT88}.  The shape
function must satisfy the following
conditions: $b(r_0)=r_0$, where $r=r_0$ is
the radius of the \emph{throat} of the
wormhole, $b'(r_0)< 1$, called the
\emph{flare-out condition}, and $b(r)<r$
for $r>r_0$.  A final requirement is
asymptotic flatness: $\text{lim}_{r\rightarrow
\infty}\nu(r)=0$ and $\text{lim}_{r\rightarrow
\infty}b(r)/r=0$.

The flare-out condition can only be met by
violating the null energy condition (NEC),
$T_{\alpha\beta}k^{\alpha}k^{\beta}\ge 0$,
for all null vectors $k^{\alpha}$, where
$T_{\alpha\beta}$ is the energy-momentum
tensor.  Matter that violates the NEC is
called ``exotic" in Ref. \cite{MT88}.  For
the outgoing null vector $(1,1,0,0)$, the
violation becomes
\begin{equation}\label{E:NEC1}
   T_{\alpha\beta}k^{\alpha}k^{\beta}=
   \rho +p_r<0.
\end{equation}
Here $T^t_{\phantom{tt}t}=-\rho$ is the
energy density, $T^r_{\phantom{rr}r}= p_r$
is the radial pressure, and
$T^\theta_{\phantom{\theta\theta}\theta}=
T^\phi_{\phantom{\phi\phi}\phi}=p_t$ is
the lateral (transverse) pressure.

Next, let us list the Einstein field
equations, referring to line element
(\ref{E:line1}):
\begin{equation}\label{E:E1}
8\pi \rho=e^{-\lambda}
\left[\frac{\lambda^\prime}{r} - \frac{1}{r^2}
\right]+\frac{1}{r^2},
\end{equation}
\begin{equation}\label{E:E2}
8\pi p_r=e^{-\lambda}
\left[\frac{1}{r^2}+\frac{\nu^\prime}{r}\right]
-\frac{1}{r^2},
\end{equation}
and
\begin{equation}\label{E:E3}
8\pi p_t=
\frac{1}{2} e^{-\lambda} \left[\frac{1}{2}(\nu^\prime)^2+
\nu^{\prime\prime} -\frac{1}{2}\lambda^\prime\nu^\prime +
\frac{1}{r}({\nu^\prime- \lambda^\prime})\right].
\end{equation}
Eq. (\ref{E:E1}) can also be written
\begin{equation}\label{E:E4}
   8\pi\rho(r)=\frac{b'(r)}{r^2}.
\end{equation}

\subsection{Fundamental problems with
   Morris-Thorne wormholes}\label{S:Problems}

The existence of exotic matter is not a
conceptual problem, as we know from the
Casimir effect \cite{MT88}.  The question
is whether enough exotic matter could be
produced to sustain a macroscopic wormhole.
According to Ref. \cite{pK02}, however,
sufficient fine-tuning might allow such a
wormhole to exist.  The origin of exotic
matter is itself an issue to be discussed
in this survey.

A violation of the NEC is a generic feature
of any traversable wormhole \cite{HV97}.
For Morris-Thorne wormholes, the violation
results from the flare-out condition.
Because of its problematical nature, it is
generally understood that the amount of
exotic matter should be kept to a minimum.
An interesting contrast is provided by
$f(R)$ modified gravity: it is possible in
principle for the throat to be threaded
with ordinary (nonexotic) matter, while the
violation of the NEC can be attributed to
the higher-order curvature terms \cite
{LO09}.  Another possibility is discussed
in Ref. \cite{pK18a}.  While the NEC is,
once again, met at the throat, the
unavoidable violation is due to the
existence of an extra spatial dimension.

A second problem to be addressed is the
enormous radial tension at the throat.
We need to recall that the radial tension
$\tau(r)$ is the negative of the radial
pressure $p_r(r)$.  It is noted in Ref.
\cite{MT88} that the Einstein field
equations can be rearranged to yield
$\tau(r)$: temporarily reintroducing
$c$ and $G$, the tension is given by
\begin{equation}
   \tau(r)=\frac{b(r)/r-[r-b(r)]\nu'(r)}
   {8\pi Gc^{-4}r^2}.
\end{equation}
It now follows that
\begin{equation}\label{E:tau}
  \tau(r_0)=\frac{1}{8\pi Gc^{-4}r_0^2}\approx
   5\times 10^{41}\frac{\text{dyn}}{\text{cm}^2}
   \left(\frac{10\,\text{m}}{r_0}\right)^2.
\end{equation}
In particular, for $r_0=3$ km, $\tau(r)$ has
the same magnitude as the pressure at the
center of a massive neutron star \cite{MT88}.
Rewriting Eq. (\ref{E:NEC1}) in the form
$\tau -\rho c^2>0$ helps explain the high
radial tension but it also ignores the
fact that exotic matter was introduced for
a completely different reason, namely
ensuring a violation of the NEC.  Furthermore,
reducing the amount of exotic matter as
much as possible makes the high radial
tension even harder to explain.  If the
amount is infinitesimal \cite{VKD},
this explanation breaks down entirely.

One can argue, of course, that the
condition $\tau-\rho c^2>0$ simply implies
that we are not dealing with ordinary
matter.  We can therefore accept the
condition and leave the details to some
advanced civilization, as suggested in
Ref. \cite{MT88}.  As noted above, however,
in $f(R)$ modified gravity, the throat
can be lined with ordinary matter, so
that the condition $\tau-\rho c^2>0$
would not be required.  But even if
$\tau<\rho c^2$, we could still be
dealing with an enormous radial tension.

Problems also occur in certain cosmological
settings.  For example, it is known that
both phantom dark energy and dark matter
can support traversable wormholes.  For
the former, the equation of state is
$p=\omega\rho$, $\omega <-1$.  Applied 
to wormholes, $\rho+p_r=\rho+\omega\rho=
(1+\omega)\rho<0$; so the NEC has been
violated.  In the case of dark matter,
the energy density is extremely low.  So
it naturally follows from Eq. (\ref{E:E4})
that $b'(r_0)=8\pi r_0^2\rho(r_0)<1$, so
that the flare-out condition is
automatically satisfied, and, just as in
the case of phantom dark energy, the NEC
is violated.  However, given the very low
energy density, Eq. (\ref{E:tau}) implies
that $r_0$ would have to be extremely
large to yield a realistic value for
$\tau(r_0)$.  On the other hand, if $r_0$
is relatively small, then the resulting
large radial tension would make little
sense and would therefore have to be
attributed, once again, to exotic matter.
But if exotic matter is needed anyway,
then both dark matter and dark energy
would become redundant.  To emphasize
this point, the zero-density case
($\rho =0$), discussed in Ref.
\cite{mVtext}, also yields a valid
wormhole solution, one that is about
as far removed from being a neutron star
as could be imagined.

The rest of this survey is devoted to
finding various ways to address these
issues.

\section{Compact stellar objects}
      \label{S:compact}
As noted in Sec. \ref{S:Problems}, a
wormhole with a relatively small throat
size could have a radial tension equal
to that of a compact stellar object such
as a neutron star.  It is shown in Ref.
\cite{pK13a} that the extreme conditions
at the center could indeed result in a
topology change, i.e., the formation of
wormholes.  The argument is based on the
fact that quark matter is likely to exist
at the center, where neutrons become
deconfined to form quark matter due
to the extreme conditions.

As a result, the analysis in Ref.
\cite{pK13a} is based on a combined
model consisting of quark matter and
baryonic matter with an isotropic
matter distribution:
\begin{equation}
  \rho_{\text{effective}}=\rho +\rho_q
  \quad \text{and} \quad
  p_{\text{effective}}=p+p_q,
\end{equation}
where $\rho$ and $p$ correspond to the
respective energy density and pressure
of the baryonic matter, while $\rho_q$
and $p_q$ correspond to the respective
energy density and pressure of the quark
matter.  (The left-hand sides refer to
the effective energy density and
pressure, respectively, of the
composition.)

In the MIT bag model, the matter
equation of state is given by
\begin{equation}
   p_q=\frac{1}{3}(\rho_q-4B),
\end{equation}
where $B$ is the bag constant.  The main
conclusion is that the topology change can
only take place if the neutron star has a
spherical core of quark matter.

Since $r=r_0$ is the throat of the wormhole,
the interior $r<r_0$ is necessarily outside
the manifold but it still contributes to
the gravitational field.  This can be
compared to a thin-shell wormhole
constructed from a Schwarzschild black
hole by the standard cut-and-paste
technique: although not part of the
manifold, the black hole generates the
gravitational field.  Moreover, the
extreme conditions, especially the
presence of quark matter, lead to
$b'(r_0)<1$, so that the flare-out
condition is satisfied.  (See Ref.
\cite{pK13a} for details.)

Since the throat of the wormhole is
deep inside the neutron star, it
cannot be directly observed.
According to Ref. \cite{DFKK},
however, indirect observations may
still be possible: if two neutron
stars are connected by a wormhole,
they would have similar
characteristics and may even exhibit
observable variations in their masses.

\section{Noncommutative geometry}
    \label{S:noncommutative}
The last several years have shown that
string theory has become ever more
influential.  An example is the
realization that coordinates may
become noncommutative operators on a
$D$-brane \cite{eW96, SW99}.  The
outcome, noncommutative geometry,
helps eliminate the divergences that
normally occur in general relativity.
The reason is that noncommutativity
replaces point-like objects by
smeared objects: spacetime can
thereby be encoded in the commutator
$[\textbf{x}^{\mu},\textbf{x}^{\nu}]
=i\theta^{\mu\nu}$, where $\theta^{\mu\nu}$
is an antisymmetric matrix that determines
the fundamental cell discretization of
spacetime in the same way that Planck's
constant $\hbar$ discretizes phase space
\cite{NSS06}.

A natural way to model the smearing effect is
by means of a Gaussian distribution of minimal
length $\sqrt{\beta}$ instead of the Dirac
delta function \cite{NSS06, NS10, mR11, fR12,
pK13b}.  An equally effective way, discussed in
Refs. \cite{pK20, LL12, NM08}, is to assume
that the energy density of the static and
spherically symmetric and particle-like
gravitational source has the form
\begin{equation}\label{E:rho}
  \rho_{\beta}(r)=\frac{\mu_1\sqrt{\beta}}
     {\pi^2(r^2+\beta)^2},
\end{equation}
where $\mu_1$ is a constant.  Eq. (\ref{E:rho})
can be interpreted to mean that the
gravitational source causes the mass $\mu_1$ of
a particle to be diffused throughout a
region of linear dimension $\sqrt{\beta}$
due to the uncertainty; so $\sqrt{\beta}$
has units of length.  Following Ref.
\cite{NSS06}, Eq. (\ref{E:rho}) leads to
the mass distribution
\begin{equation}
   m(r)=\int^r_04\pi(r')^2\rho(r')dr'
   =\frac{2M}{\pi}\left(\text{tan}^{-1}
   \frac{r}{\sqrt{\beta}}-
   \frac{r\sqrt{\beta}}{r^2+\beta}\right),
\end{equation}
where $M$ is now the total mass of the
source.  Observe that $m(0)=0$ but $m(r)$
rapidly rises to $M$ as $r$ increases.

According to Ref. \cite{NSS06}, noncommutative
geometry is an intrinsic property of spacetime
and does not depend on any particular features
such as curvature.  Moreover, the relationship
between the radial pressure and energy density
is given by
\begin{equation}\label{E:pressure}
  p_r=-\rho_{\beta}.
\end{equation}
The reason is that the source is a
self-gravitating droplet of anisotropic
fluid of density $\rho_{\beta}$ and the
radial pressure is needed to prevent a
collapse to a matter point.  From Eqs.
(\ref{E:rho}) and (\ref{E:E4}), we now
obtain
\begin{multline}\label{E:shape}
  b(r)=\frac{8M\sqrt{\beta}}{\pi}\int^r_{r_0}
  \frac{(r')^2dr'}{[(r')^2+\beta]^2}+r_0\\
  =\frac{4M\sqrt{\beta}}{\pi}
  \left(\frac{1}{\sqrt{\beta}}\text{tan}^{-1}
  \frac{r}{\sqrt{\beta}}-\frac{r}{r^2+\beta}
  -\frac{1}{\sqrt{\beta}}\text{tan}^{-1}
  \frac{r_0}{\sqrt{\beta}}+\frac{r_0}{r_0^2
  +\beta}\right)+r_0.
\end{multline}
It is shown in Ref. \cite{pK15} that
$b=b(r)$ has the usual properties of a
shape function: $b(r_0)=r_0$, $0<b'(r_0)
<1$, and $\text{lim}_{r\rightarrow
\infty}b(r)/r=0$.

For the purpose of this survey, the
most important formula is the density
$\rho_s$ of the surface of the throat,
\begin{equation}\label{E:surface}
   \rho_s=\frac{\mu_2\sqrt{\beta}}{\pi^2
   [(r-r_0)^2+\beta]^2},
\end{equation}
where $\mu_2$ is the mass of the surface.
As noted in Sec. \ref{S:Problems}, we are
dealing with a moderate throat size.
Also, the surface $r=r_0$ is a smeared
surface since the individual particles
on the surface are smeared.  (For further
discussion, see Ref. \cite{pK20}.)

Unlike the case of dark matter, a low
energy density does not result in a low
tension \cite{pK20}.  Recalling that the
tension $\tau$ is the negative of the
pressure $p$, $p_r+\rho_s<0$ becomes
$\tau-\rho_s>0$.  Eq. (\ref{E:pressure})
implies that we are right on the edge of
violating the NEC, i.e., $\tau-\rho_s=0$.
So at $r=r_0$,
\begin{equation}\label{E:high}
  \rho_s=\frac{\mu_2}{\pi^2}\frac{1}{\beta^{3/2}},
\end{equation}
but we still have $\tau-\rho_s=0$.  Since
the throat is a smeared surface, however,
we only have $r\approx r_0$.  So by Eq.
(\ref{E:surface}), $\rho_s$ is reduced in
value and we obtain the desired
$\tau-\rho_s>0$.   Eq. (\ref{E:high}) implies
that $\rho_s$ and $\tau$ are extremely large
at the throat.  To check its plausibility,
it is argued in Ref. \cite{pK20} that,
according to Eq. (\ref{E:tau}), for a
throat size of 10 m,
$\tau\approx 5\times 10^{41} \,\text{dyn}
/\text{cm}^2$. Suppose $\mu_2$ has the
rather minute value $10^{-10}$ g.  Applying
Eq. (\ref{E:high}), we have
\begin{equation}
  \tau=\rho_sc^2=\frac{\mu_2}{\pi^2}
  (\sqrt{\beta})^{-3}c^2=5\times 10^{41}
  \,\frac{\text{dyn}}{\text{cm}^2}.
\end{equation}
To satisfy this relationship, the value
$\sqrt{\beta}=10^{-11}\,\text{cm}$ is
sufficient.  Since $\sqrt{\beta}$ may be
much smaller, we can accommodate even
larger values of $\tau$.

We conclude that even though we are
dealing with a very low energy density,
we have a very large tension at the throat
thanks to the noncommutative-geometry
background.

\section{Returning to $f(R)$ modified
    gravity}\label{S:modified}

Modified gravitational theories, including
$f(R)$ modified gravity, have been invoked
to explain various phenomena in general
relativity.  First we need to recall that
in $f(R)$ gravity, the Ricci scalar $R$
in the Einstein-Hilbert action $S_{\text{EH}}
=\int\sqrt{-g}\,R\,dx^4$ is replaced by
a nonlinear function $f(R)$.  Next,
let us state the gravitational field
equations in the form used by Lobo and
Oliveira \cite{LO09}:
\begin{equation}\label{E:Lobo1}
   \rho(r)=F(r)\frac{b'(r)}{r^2},
\end{equation}
\begin{equation}\label{E:Lobo2}
   p_r(r)=-F(r)\frac{b(r)}{r^3}+F'(r)
   \frac{rb'(r)-b(r)}{2r^2}-F''(r)
   \left[1-\frac{b(r)}{r}\right],
\end{equation}
and
\begin{equation}\label{E:Lobo3}
   p_t(r)=-\frac{F'(r)}{r}\left[1-
   \frac{b(r)}{r}\right]+\frac{F(r)}
   {2r^3}[b(r)-rb'(r)];
\end{equation}
here $F=\frac{df}{dR}$.  It is also
assumed that $\nu(r)\equiv\text{constant}$,
so that $\nu'(r)\equiv 0$.  Otherwise,
according to Ref. \cite{LO09}, the
analysis becomes intractable.

According to Ref. \cite{pK20}, by assuming
a noncommutative-geometry background, we
obtain in view of Eq. (\ref{E:rho}),
\begin{equation}\label{E:f(R)}
    f(R)=
    \frac{\mu_1\sqrt{\beta}}{\pi^2}
    \frac{(\beta R+2\alpha)\text{ln}\,
    (\beta R+2\alpha)-\beta R}
    {\beta^2(\beta R+2\alpha)}+C,
\end{equation}
where $C$ and $\alpha$ are constants,
thereby providing a motivation for
the choice of $f(R)$.

From Eq. (\ref{E:Lobo1}),
\begin{equation}
   b'(r_0)=\frac{r_0^2\rho(r_0)}
      {F(r_0)}<1,
\end{equation}
even if $F(r_0)$ is quite small, thereby
satisfying the flare-out condition.
Next, from Eq. (\ref{E:Lobo2}),
\begin{equation}
   \tau(r_0)=-p_r(r_0)=F(r_0)\frac{b(r_0)}
   {r_0^3}-F'(r_0)\frac{r_0b'(r_0)-b(r_0)}
   {2r_0^2}.
\end{equation}
So given that $b'(r_0)<1$, $\tau(r_0)$
is large provided that $F'(r_0)$ is
large and positive.  It is shown in
Ref. \cite{pK20} that this requirement
can be met if we retain the connection
to noncommutative geometry via $f(R)$
in Eq. (\ref{E:f(R)}).

Unlike the situation in classical
general relativity, meeting the
flare-out condition does not
automatically lead to a violation
of the NEC, but, as noted in Sec.
\ref{S:Problems}, the unavoidable
violation can be attributed to the
higher-order curvature terms
\cite{LO09}.  For this conclusion
to hold, noncommutative geometry
must be viewed as a special case of
$f(R)$ modified gravity.  The
connection between the two theories
had already been suggested in Ref.
\cite{pK18b}.

\section{A small extra spatial
   dimension}\label{S:extra}
   It was brought out in Sec.
\ref{S:Problems} that the throat of
a wormhole could be threaded with
ordinary (nonexotic) matter if the
violation of the NEC can be attributed
to either the higher-order curvature
terms in $f(R)$ modified gravity
\cite{LO09}\emph{ }or to the existence
of a higher spatial dimension
\cite{pK18a}.  The latter case is
based on the line element
\begin{equation}\label{E:L4}
  ds^2=-e^{2\Phi(r)}dt^2
  +\frac{dr^2}{1-b(r)/r}+r^2
  (d\theta^2+\text{sin}^2\theta\,d\phi^2)
  +e^{2\mu(r,l)}dl^2,
\end{equation}
where $l$ is the extra spatial dimension.
To study the radial tension, Ref.
\cite{pK20} starts with the
Einstein field equations in the
orthonormal frame:
\begin{equation}
   G_{\hat{\alpha}\hat{\beta}}=
   R_{\hat{\alpha}\hat{\beta}}-\frac{1}{2}R
   g_{\hat{\alpha}\hat{\beta}}
   =\kappa T_{\hat{\alpha}\hat{\beta}},
\end{equation}   
where $\kappa$ is the coupling constant.  
(In the five-dimensional case, $\kappa=3\pi^2$.)   
Also,
\begin{equation}
   g_{\hat{\alpha}\hat{\beta}}=
   \left(
   \begin{matrix}
   -1&0&0&0&0\\
   \phantom{-}0&1&0&0&0\\
   \phantom{-}0&0&1&0&0\\
   \phantom{-}0&0&0&1&0\\
   \phantom{-}0&0&0&0&1
   \end{matrix}
   \right)
\end{equation}
and $\tau(r)=-p_r(r)=-T_{11}$.  The
components of the Ricci scalar are
given in Ref. \cite{pK20}.  From
$G_{11}=\kappa T_{11}$, we obtain
\begin{equation}
   \kappa p_r(r)=R_{11}-\frac{1}{2}Rg_{11},
\end{equation}
while
\begin{equation}\label{E:Ricci}
R=R^i_{\phantom{0}i}=-R_{00}+R_{11}
  +R_{22}+R_{33}+R_{44}.
\end{equation}
It follows that
\begin{equation}
   2\kappa p_r(r)=R_{00}+R_{11}
  -R_{22}-R_{33}-R_{44}.
\end{equation}
The fifth dimension comes into play
because (Ref. \cite{pK20})
\begin{equation}\label{E:radial}
   2\kappa p_r(r)=-\frac{1}{2r^2}\frac
   {\partial\mu(r,l)}{\partial r}
   \left[rb'(r)-b(r)\right]-
      \frac{2b(r)}{r^3},
\end{equation}
which is negative if $\partial\mu(r,l)
/\partial r<0$.  Eq. (\ref{E:radial})
is a tension whose magnitude depends on
$\partial\mu(r,l)/\partial r$.  In
principle, then, $\tau(r)$ can be
extremely large.

Finally, according to Ref. \cite{pK20},
the extra dimension can be small or
even curled up, as in string theory.
In fact, the models discussed in
Sections \ref{S:noncommutative},
\ref{S:modified}, and \ref{S:extra}
help confirm that whenever we are
dealing with an extreme regime, the
effects of string theory cannot be
neglected \cite{lS93}.

\section{Solutions based on the theory of
   embedding}\label{S:timespace}
Using embedding theorems to account for
the high radial tension has the advantage
of offering a possible explanation for
the origin of exotic matter.  Embedding
theorems have a long history in the
general theory of relativity, in large
part due to Campbell's theorem \cite{jC26}.
According to Ref. \cite{pW15}, the field
equations in terms of the Ricci scalar
are $R_{AB}=0$, $A, B=0, 1, 2, 3, 4$.
The resulting five-dimensional theory
explains the origin of matter in the
following sense: the vacuum field
equations in five dimensions yield
the usual Einstein field equations
\emph{with matter}, called the
induced-matter theory \cite{PW92,
SW03}.  What we perceive as matter
is the impingement of the fifth
dimension onto our spacetime -- and
this would include exotic matter.
It even suggests that the amount of
exotic matter may be irrelevant.
Given its problematical nature,
however, we would prefer that the
amount be kept to a minimum.
Fortunately, the embedding theory
fulfills this requirement.

According to Campbell's theorem
\cite{jC26}, a Riemannian space can
be embedded in a higher-dimensional
flat space: an $n$-dimensional
Riemannian space is said to be of
embedding class $m$ if $m+n$ is the
lowest dimension $d$ of the flat space
in which the given space can be
embedded; here $d=\frac{1}{2}n(n-1)$.
So a four-dimensional Riemannian
space is of class two and can
therefore be embedded in a
six-dimensional flat space, i.e., $d=6$.
Moreover, a line element of class
two can be reduced to a line
element of class one by a suitable
coordinate transformation
\cite{MDRK, MRG, MM17, NG17}.

An interesting aspect of Einstein's
theory is that the extra dimension
can be spacelike or timelike.  These
cases will be taken up separately.

\subsection{An extra timelike dimension}
    \label{S:time}

Due to the extra timelike dimension,
the embedding space has the form
\begin{equation}\label{E:line3}
   ds^2=-(dz^1)^2-(dz^2)^2+(dz^3)^2+
   (dz^4)^2+(dz^5)^2.
\end{equation}
According to Ref. \cite{pK21a},
the coordinate transformation is
$z^1=\sqrt{K}\,e^{\nu/2}\,\text{sin}
\frac{t}{\sqrt{K}}$, $z^2=
\sqrt{K}\,e^{\nu/2}\,\text{cos}
\frac{t}{\sqrt{K}}$,
$z^3=r\,\text{sin}\,\theta\,
\text{cos}\,\phi$, $z^4=r\,\text{sin}
\,\theta\, \text{sin}\,\phi$, and
$z^5=r\,\text{cos}\,\theta$.
Substituting in Eq. (\ref{E:line3})
yields the line element
\begin{equation}\label{E:line4}
ds^{2}=-e^{\nu}dt^{2}
 +\left[1-\frac{1}{4}Ke^{\nu}(\nu')^2\right]dr^2
+r^{2}(d\theta^{2}+\text{sin}^{2}\theta\,
d\phi^{2}).
\end{equation}
This metric is equivalent to
metric (\ref{E:line1}) if
\begin{equation}\label{E:lambda2}
   e^{\lambda}=1-\frac{1}{4}Ke^{\nu}(\nu')^2,
\end{equation}
where $K>0$ is a free parameter.

Given Eq. (\ref{E:lambda2}), Eq. (\ref{E:E2})
immediately yields
\begin{equation}
   p_r(r)=\frac{1}{8\pi}\left[
   \frac{1}{1-\frac{1}{4}Ke^{\nu}(\nu')^2}
   \left(\frac{1}{r^2}+\frac{\nu'}{r}
   \right)-\frac{1}{r^2}\right].
\end{equation}
Since $K$ is a free parameter, $K$ can
be chosen to make $\tau(r)=-p_r(r)$ as
large as required without relying on
exotic matter.  It turns out, however,
that to meet the other conditions, some
fine-tuning cannot be avoided.  For
example, Eq. (\ref{E:lambda2}) yields
\begin{equation}
   b(r)=r\left[1-\frac{1}
   {1-\frac{1}{4}Ke^{\nu(r)}[\nu'(r)]^2}
   \right]+\frac{r_0}{1-\frac{1}
   {4}Ke^{\nu(r_0)}[\nu'(r_0)]^2},
\end{equation}
so that $b(r_0)=r_0$, but the flare-out
condition can only be met if $(\nu')^2
+2\nu''$ is sufficiently small near the
throat.  (See Ref. \cite{pK21a} for
details.)  Fortunately, this condition
can be easily met ``by hand.''  The
same condition ensures that $\rho+p_r$
can be made as small as desired.  So
the amount of exotic matter \cite{NZK},
\begin{equation}\label{E:Nandi}
   \Omega =\int^{2\pi}_0\int^{\pi}_0
   \int^{\infty}_{r_0}(\rho +p_r)
   \sqrt{-g}\,\,drd\theta d\phi,
\end{equation}
can also be kept small.

Finally, the resulting wormhole spacetime
is asymptotically flat.

\subsection{An extra spacelike dimension}
     \label{S:space}

If the extra dimension is spacelike,
then the embedding space has the form
\begin{equation}\label{E:line4}
   ds^2=-(dz^1)^2+(dz^2)^2+(dz^3)^2+
   (dz^4)^2+(dz^5)^2.
\end{equation}
This case is discussed in Ref.
\cite{pK21b}.  Here the coordinate
transformation is \cite{MDRK, MRG}
$z^1=\sqrt{K}\,e^{\nu/2}\,\text{sinh}
\frac{t}{\sqrt{K}}$, $z^2=
\sqrt{K}\,e^{\nu/2}\,\text{cosh}
\frac{t}{\sqrt{K}}$,
$z^3=r\,\text{sin}\,\theta\,
\text{cos}\,\phi$, $z^4=r\,\text{sin}
\,\theta\, \text{sin}\,\phi$, and
$z^5=r\,\text{cos}\,\theta$.  This
time we obtain
\begin{equation}
ds^{2}=-e^{\nu}dt^{2}
 +\left[1+\frac{1}{4}Ke^{\nu}(\nu')^2\right]dr^2
+r^{2}(d\theta^{2}+\text{sin}^{2}\theta\,
d\phi^{2})
\end{equation}
and
\begin{equation}\label{E:lambda3}
   e^{\lambda}=1+\frac{1}{4}Ke^{\nu}(\nu')^2.
\end{equation}
The result is a metric of embedding class
one.  Eq. {(\ref{E:lambda3}) can also be
obtained from the Karmarkar condition
\begin{equation*}
  R_{1414}=
  \frac{R_{1212}R_{3434}+R_{1224}R_{1334}}
  {R_{2323}},\quad R_{2323}\neq 0.
\end{equation*}
In fact, Eq. (\ref{E:lambda3}) is a solution
of the differential equation
\begin{equation*}
   \frac{\nu'\lambda'}{1-e^{\lambda}}=
   \nu'\lambda'-2\nu''-(\nu')^2,
\end{equation*}
making $K$ an arbitrary constant of
integration.

To continue, Ref. \cite{pK21b} makes
use of the following redshift function
first proposed by Lake \cite{kL03}:
\begin{equation}\label{E:Lake}
   \nu(r)=n\,\text{ln}\,(1+Ar^2),\quad
   n\geq 1,
\end{equation}
where $A$ is a constant.  According
to Ref. \cite{kL03}, this class of
monotone increasing functions generates
all regular static spherically symmetric
perfect-fluid solutions of the Einstein
field equations.  Eq. (\ref{E:Lake})
can be written $e^{\nu}=(1+Ar^2)^n$.
A slightly more general form,
$e^{\nu}=B(1+Ar^2)^n$, is also
acceptable since the resulting $\nu$
is still monotone increasing.  For
convenience, Ref. \cite{pK21b}
assumes that $n=1$.  The resulting
form is
\begin{equation}
   e^{\nu}=B(1+Ar^2),\quad A,B>0.
\end{equation}
While $A$ is still a free parameter,
$B$ can be determined from the
junction conditions discussed below.

This time the shape function is
\begin{equation}\label{E:shape}
   b(r)=r\left[1-\frac{1}
   {1+\frac{1}{4}Ke^{\nu(r)}[\nu'(r)]^2}
   \right]+\frac{r_0}{1+\frac{1}
   {4}Ke^{\nu(r_0)}[\nu'(r_0)]^2}.
\end{equation}
It is shown in Ref. \cite{pK21b} that
for sufficiently large $K$,
\begin{equation}\label{E:bprime2}
   b'(r_0)=1+\frac{1-Ar_0^2}
      {1+Ar_0^2+KBA^2r_0^2}.
\end{equation}\emph{\emph{\emph{}}}
It follows that
\begin{equation}\label{E:flair}
   b'(r_0)<1 \quad \text{for} \quad
   r_0>\frac{!}{\sqrt{A}}.
\end{equation}
So the flare-out condition is met
whenever $r_0>1/\sqrt{A}$.

While $\text{lim}_{r\rightarrow
\infty}b(r)/r=0$, it is not true that
$\text{lim}_{r\rightarrow\infty}
\nu(r)=0$.  So the wormhole spacetime
is not asymptotically flat and must
therefore be cut off at some $r=a$ and
joined to an exterior Schwarzschild
spacetime.  In other words,
$e^{\nu(a)}=B(1+Aa^2)=1-2M/a$;  so
\begin{equation}
   B=\frac{1-\frac{2M}{a}}{1+Aa^2}.
\end{equation}

For the radial pressure, we get
\begin{equation}
   p_r(r)=\frac{A(2-KBA)}{8\pi(1+Ar^2
      +KBA^2r^2)},
\end{equation}
whence
\begin{equation}\label{E:limittau}
  \text{lim}_{r_0\rightarrow 0} \tau(r_0)
  =\frac{A}{8\pi}(-2+KBA).
\end{equation}
So $\tau(r_0)$ can be made as large as
required due to the free parameter $K$,
as long as $r_0$ is relatively small.

Finally, according to Ref. \cite{pK21b},
 $\text{lim}_{K\rightarrow \infty}
 8\pi(\rho+p_r)=0$, showing that the
 amount of exotic matter can be kept
 to a minimum.

 \section{Could the Universe be
   multiply connected?}\label{S:donut}

A commonly accepted view is that
the Universe is a 3-sphere and hence
simply connected.  The idea of a
multiply-connected Universe has
recently been revived \cite{rA21}.
We saw in Sec. \ref{S:Problems} that
this possibility is a direct
consequence of Einstein's theory in
the following sense: if we assume a
dark-matter or dark-energy background,
the conditions for the existence of
wormholes are satisfied, but such
wormholes could only exist on very
large scales.  An example of such a
wormhole would be the huge doughnut
shape proposed in Ref. \cite{rA21}.

\section{Summary}
Wormholes are just as good a prediction
of Einstein's theory as black holes, in
part because Einstein's theory is able
to tolerate strange-sounding features
such as space and time warps.  Quantum
field theory, on the other hand, makes
its own demands that cannot be readily
dismissed.  For example, the need to
violate the null energy condition (NEC)
is a generic feature of a traversable
wormhole, calling for the existence
of exotic matter.  Not only that,
exotic matter must exist in sufficient
quantities, which, on a macroscopic
scale, is itself problematical.
Another serious issue is the enormous
radial tension at the throat, which
could exceed the tension at the center
of a massive neutron star unless the
throat radius is extremely large.
Attributing this characteristic of
exotic matter ignores the fact that
exotic matter was introduced for a
completely different reason.  Reducing
the amount thereof, by itself highly
desirable, makes matters worse.  If
the amount is infinitesimal, then
exotic matter could not have the
slightest effect on the radial
tension.  In $f(R)$ modified gravity,
the throat can be lined with ordinary
(nonexotic) matter, but as noted in
Sec. \ref{S:Problems}, this does not
necessarily avoid a high radial
tension.

These issues are addressed in four
recent publications \cite{pK13a,
pK20, pK21a, pK21b}, covering six
different aspects.  Sec.
\ref{S:compact}, which is based on
Ref. \cite{pK13a}, deals with neutron
star interiors.  According to Ref.
\cite{pK13a}, the large radial
tension is sufficient for a topology
change, i.e., the formation of
wormholes, provided that there is a
core of quark matter at the center.

Sections \ref{S:noncommutative},
\ref{S:modified}, and \ref{S:extra}
are based on Ref. \cite{pK20}.  Sec.
\ref{S:noncommutative} briefly
discusses wormholes in a
noncommutative-geometry setting.
The low energy density causes the
flare-out condition to be met,
thereby resulting in a violation
of the NEC at or near the throat.
While this is also true of wormholes
supported by dark matter and dark
energy, it is only the
noncommutative-geometry case that
results in a large radial tension.
Since the NEC has been duely
violated, the large radial tension
can indeed be attributed to exotic
matter.  Sec. \ref{S:modified}
combines the noncommutative-geometry
background with $f(R)$ modified
gravity by deriving the form of
$f(R)$, Eq. (\ref{E:f(R)}).  In
this way the large radial tension
is retained.  The throat can now
be threaded with ordinary (nonexotic)
matter, however, since the
violation of the NEC, the generic
feature mentioned earlier, can be
attributed to the higher-order
curvature terms in the modified
theory.  Conceptually, this is a
considerable advantage since the
large radial tension is not
directly connected to exotic matter,

Sec. \ref{S:extra} assumes an extra
spatial dimension, possibly small
or even curled up.  It is noted that,
once again, the throat could be lined
with ordinary matter because the
violation of the NEC can be attributed
to the extra spatial dimension.  It
is shown in Ref. \cite{pK20} that
the extra dimension can also account
for the large radial tension.

Sec. \ref{S:timespace} summarizes the
theory of embedding, including the
induced-matter theory: what we
perceive as matter is the impingement
of the higher-dimensional space onto
ours.  This may include exotic matter,
suggesting that the amount of exotic
matter is irrelevant.  Given its
problematical nature, however, it is
generally assumed that the amount
should be kept to a minimum.  The
embedding theory can fulfill both
of these requirements.

A Riemannian space can be embedded
in a higher-dimensional flat space:
an $n$-dimensional Riemannian space
is said to be of embedding class
$m$ if $m+n$ is the lowest dimension
of the flat space in which the given
space can be embedded.  In particular,
a four-dimensional Riemammian space
is of class two but can be reduced
to class one by a suitable
transformation of coordinates.
Furthermore, Einstein's theory
allows the extra dimension to be
either timelike or spacelike.  The
timelike case is discussed in Sec.
\ref{S:time}, based on Ref.
\cite{pK21a}, and the spacelike
case is discussed in Sec.
\ref{S:space}, based on
Ref. \cite{pK21b}.  In both cases,
the free parameter $K$ in the
coordinate transformation can be
chosen to make the radial tension \
as large as required without
relying on exotic matter.

Finally, Sec. \ref{S:donut} discusses
the possibility of a multiply-connected
Universe.

\end{document}